\documentclass[pre,aps,twocolumn,eqsecnum]{revtex4}
\usepackage{epsfig}
\usepackage{bm}
\usepackage{lscape}

\newcommand{\C}[1]{{\mathcal{#1}}}
\usepackage[latin1]{inputenc}
\newcommand{\beq}{\begin{equation}}
\newcommand{\eeq}{\end{equation}}
\newcommand{\bea}{\begin{eqnarray}}
\newcommand{\eea}{\end{eqnarray}}

\begin{document}

\title{Athermal Shear-Transformation-Zone Theory of Amorphous Plastic Deformation II: Analysis of Simulated Amorphous Silicon}

\author{Eran Bouchbinder$^1$, J. S. Langer$^2$, and Itamar Procaccia$^1$}
\affiliation{$^1$ Dept. of Chemical Physics, The Weizmann Institute
of Science, Rehovot 76100 Israel.\\
$^2$ Dept. of Physics, University of California, Santa
Barbara, CA  93106-9530  USA}

\date{\today}

\begin{abstract}
In the preceding paper, we developed an athermal shear-transformation-zone
(STZ) theory of amorphous plasticity. Here we use this theory in an 
analysis of  numerical simulations of plasticity in amorphous silicon by
Demkowicz and Argon (DA). In addition to bulk mechanical properties,
those authors observed internal features of their deforming
system that challenge our theory in important ways. 
We propose a quasithermodynamic interpretation of their observations 
in which the effective disorder temperature, generated by mechanical deformation 
well below the glass temperature, governs the behavior of other state 
variables that fall in and out of equilibrium with it. 
Our analysis points to a limitation of either the step-strain procedure 
used by DA in their simulations, or the STZ theory in its 
ability to describe rapid transients in stress-strain curves, or 
perhaps to both. Once we allow for this limitation, we are able to 
bring our theoretical predictions into accurate agreement with the simulations.

\end{abstract}

\maketitle

\section{Introduction}

In the preceding paper \cite{BLP06}, we presented an athermal STZ theory 
of plastic deformation in materials where thermal activation of 
irreversible molecular rearrangements is negligible or nonexistent.  
Here we use that theory to interpret an extensive body of 
computational data published recently by Demkowicz and Argon \cite{04DA,DA1,DA2,06AD}, 
hereafter referred to occasionally as ``DA.'' 
Those authors simulated plastic deformation in amorphous 
silicon using a system of 4096 atoms
interacting {\it via} a Stillinger-Weber potential \cite{85SW} in a
cubic cell with periodic boundary conditions.  They subjected this
system to pure shear under both constant-volume and
constant-zero-pressure, plane-strain, conditions. Their 
reports of these simulations are remarkably complete and 
detailed.  They provide valuable 
and challenging information about: 
\begin{itemize}
\item the relation between stress-strain response and sample preparation;
\item the theoretical description of nonequilibrium behavior in systems subject to steady-state and transient mechanical deformation;
\item the nature of the glass transition in simulated amorphous silicon;
\item the strengths and limitations of numerical simulation techniques.
\end{itemize}
We address each of these topics in the body of this report.

Demkowicz and Argon used two different procedures for
simulating shear deformation.  In their potential energy 
minimization method (PEM), each step in the process consisted 
of a small, affine, shear displacement
of all the atomic positions, followed by a minimization of the
potential energy during which the atoms relax to their nearby positions of
mechanical equilibrium.   Supposedly, PEM simulations correspond to
the limit of zero strain rate at zero temperature, but that 
interpretation is problematic. In their molecular 
dynamics (MD) method, DA used a different step-strain procedure in 
which each small shear increment was followed by an MD 
relaxation at temperature $T = 300\,K$, with an average 
strain rate of order $10^8\,{\rm sec}^{-1}$. In both procedures, an
incremental shear was imposed only after the system was judged to
have reached a stable, stationary state following the preceding
step.  We argue below that there are important uncertainties associated 
with both of these step-strain simulation methods.  

The single most important feature of the DA simulations is
that, in addition to measuring the shear stress (and keeping track
of pressure and/or volume changes) during deformation, DA also
observed local atomic correlations within their system.  Here they
were taking advantage of their numerical method to see inside their
system in a way that is seldom possible in laboratory experiments
using real materials. They found that the environments of some atoms
were solidlike and others liquidlike, and that the liquidlike
regions seemed to be, as they say, the ``plasticity carriers.''
Before any mechanical deformation, their fraction of liquidlike
regions $\phi$ was small when the system was annealed or cooled slowly, and
was approximately 0.5 when the system was quenched
rapidly.  Then, during  constant-zero-pressure deformations, $\phi$
approached a value slightly less than 0.5, independent of its
initial value.  Thus, $\phi$ behaved in a manner qualitatively
similar to the dimensionless effective temperature $\chi$ or, equivalently, the
density of STZ's $\Lambda \propto \exp (-1/\chi)$ described in 
\cite{BLP06}.  The relation between these quantities is one of the main 
topics to be addressed in this paper.

Amorphous silicon, like water, expands as it solidifies.  Moreover,
its properties are highly sensitive to small changes in density.  A
slowly quenched, more nearly equilibrated system is less dense than
one that is rapidly quenched, because the former contains a smaller
population of denser, liquidlike regions than the latter.  When strained, the
more nearly equilibrated system initially responds elastically, and
takes longer than a rapidly quenched system to generate enough
plasticity carriers to enable plastic flow.  Accordingly, the
equilibrated system exhibits a more pronounced stress peak of the
kind illustrated in Figure 1 (top curve, upper panel) of \cite{BLP06}.  
As the liquidlike fraction $\phi$ increases, the system contracts 
if held at constant pressure; or else, if the system is held 
at constant volume, the pressure decreases and may even become 
negative.  It may be surprising to theorists but nevertheless is
true that, in this material, the STZ's must be associated with 
decreases rather than increases in free volume.

For theoretical purposes, it will be simplest to look only at
constant-pressure simulations, because the changes in pressure that
occur at constant volume require additional, hard-to-control
approximations for behaviors that are not intrinsic to the STZ
hypotheses.  System-specific parameters such as the yield stress $s_y$ 
are likely to be more sensitive to changes in
pressure than in volume.  For example, the Mohr-Coulomb effect
implies that $s_y$ increases with pressure. Demkowicz and Argon 
point out that direct evidence for
pressure dependence can be seen in the graphs of the liquid fraction
$\phi$ as a function of strain $\gamma$ in Figs. 10(c) and 15(c) of
\cite{DA1}.  In the constant-zero-pressure simulations shown in their Fig.
15, $\phi$ equilibrates to approximately the same value after
deformation for all four different initial conditions.  That does
not happen in the constant-volume simulations in Fig. 10, implying
that the internal states of these deformed systems differ from one
another in nontrivial ways. 

In addition to the questions pertaining to step-strain procedures,  
to be discussed in detail in Sec. \ref{SecondThoughts}, there are other 
limitations of the DA simulations that must be taken into account.
The DA simulation system is too small to make it likely that more 
than one STZ-triggered event is taking place within a characteristic 
plastic relaxation time; and data is reported only for single 
numerical experiments performed on individual samples rather than 
averages over multiple experiments.  Thus, fluctuations are large, 
and the results must be sensitive to statistical variations 
in initial conditions.  In short, we must be careful in our interpretations.

On the positive side, Demkowicz and Argon's remarkable
combination of multiple simulation methods and multiple 
observations allows us to use the athermal STZ theory 
to construct what we believe is an internally self-consistent 
interpretation of their data.  Going beyond their stress-strain 
curves, and focusing on the behaviors of the liquidlike 
fraction $\phi$ and the density $\rho$, we develop a quasithermodynamic 
picture in which the effective disorder temperature $\chi$ plays
a dominant role. In this picture, the STZ's, {\it i.e.} the active flow defects, 
are rare sites, out in the wings of the disorder distribution, 
that are more susceptible than their neighbors to 
stress-induced shear transformations.  They almost certainly 
lie in the liquidlike regions of the system, but only a 
very small fraction of the liquidlike sites are STZ's. The energy 
dissipated in the STZ-initiated transitions generates the effective 
temperature $\chi$, a systemwide intensive quantity that, 
in turn, determines extensive quantities such as 
$\phi$ and $\rho$. 

Our central hypothesis is that there exist 
quasithermodynamic equations of state relating steady-state values of 
$\phi$ and $\rho$ to $\chi$ (and also, in principle, to the pressure and shear stress).
Once we have found these equations of state, we extend the 
quasithermodynamic model to describe the way in which  $\phi$ and $\rho$
fall in and out of equilibrium with $\chi$ during transient 
responses to external driving forces. In this way, we arrive at a 
quantitative interpretation of the DA simulations.   

We begin in Sec. \ref{STZsummary} with a brief summary of the athermal
STZ theory developed in \cite{BLP06}.  Section \ref{FirstFits} contains 
a preliminary analysis of the DA stress-strain data. In Sec. \ref{SecondThoughts},
we explain why we cannot accept the initially good agreement between the STZ theory
and the simulation data.  These arguments motivate the quasithermodynamic 
hypothesis, introduced in Sec. \ref{Quasithermodynamics}.  We extend the 
quasithermodynamic ideas to nonequilibrium situations in Sec. \ref{Departures}. 
Finally, in Sec. \ref{Conclusions}, we conclude with some speculations concerning 
the validity of the quasithermodynamic picture and its relation to the STZ theory in general.

\section{STZ Summary}
\label{STZsummary}

In order to make this paper reasonably self-contained, 
we start by restating Eqs. (3.12) - (3.16) in \cite{BLP06}.  
The first of these equations is an expression for the 
total rate of deformation tensor $D^{tot}= \dot\gamma/2$ 
as the sum of elastic and plastic parts:
\begin{equation}
\label{Dtot}
D^{tot} = {\dot{\tilde s}\over 2\,\tilde\mu} + D^{pl}(\tilde s, m, \Lambda),
\end{equation}
where $\tilde s$ and $\tilde\mu$ are the deviatoric stress and the shear modulus measured in units of the yield stress $s_y$.  The internal state variables $m$ and $\Lambda$ are, respectively, the orientational bias and the scaled density of STZ's. We consider only the case of pure, plane-strain shear in which the material is strained at a fixed rate $\dot\gamma$, and the stress is measured as a function of the strain $\gamma$.  To describe such experiments, we write Eq.(\ref{Dtot}) in the form
\begin{equation}
\label{sgamma}
{d \tilde s\over d\gamma}= \tilde\mu\,\left(1 - {2\,\epsilon_0\,\Lambda\over \dot\gamma\,\tau_0}\,q(\tilde s, m)\right),
\end{equation}
where $\epsilon_0$ is a number of order unity (the atomic density in atomic units multiplied by the incremental strain associated with an STZ transition), $\tau_0$ is the characteristic time scale for STZ transitions, and 
\begin{equation}
\label{qdef}
q(\tilde s,m) \equiv \C C(\tilde s)\,\Bigl({\tilde s\over |\tilde s|}-m\Bigr).
\end{equation}
The function ${\cal C}(\tilde s)$ describes the stress dependence of the STZ transition rate; it is proportional to $|\tilde s -1|$ for $|\tilde s|\gg 1$ and vanishes smoothly near $\tilde s = 0$.  As defined by Eq.(3.6) in \cite{BLP06}, the smoothness of ${\cal C}(\tilde s)$ near $\tilde s=0$ is determined by a parameter $\zeta$, which we take to be unity throughout the following.  

Similarly, we  restate Eqs. (3.9), (3.10) and (3.11) from \cite{BLP06}:
\begin{equation}
\label{dotm4}
{dm\over d\gamma} = {2\over \dot\gamma\,\tau_0}\,q(\tilde s,m)\,\Bigl(1 - {m\,\tilde s\over \Lambda}\,e^{-1/\chi}\Bigr);
\end{equation}
\begin{equation}
\label{dotLambda4}
{d\Lambda\over d\gamma} = {2\over\dot\gamma\,\tau_0}\,\tilde s\,q(\tilde s,m)\,\bigl(e^{-1/\chi} - \Lambda\bigr);
\end{equation}
and
\begin{equation}
\label{dotchi4}
{d\chi\over d\gamma} = {2\,\epsilon_0\over c_0\,\dot\gamma\,\tau_0}\,\Lambda\,\tilde s\,q(\tilde s,m)\,(\chi_{\infty}-\chi).
\end{equation}
The constant $c_0$ is a configurational specific heat per atom in units $k_B$, which must be  of order unity.

We assume that the stress and strain
tensors in these equations remain diagonal in two dimensional pure
shear and plane strain, and that any correction in the third
dimension is negligible in comparison to other uncertainties in this
analysis.  A basic premise of the STZ theory is that the zones are rare and do not interact with each other; thus we assume from
the beginning that the quantity $\epsilon_0\,\exp(-1/\chi)$ is small
of order $10^{-3}$ (in fact, very much less), so that the equations for $m(\gamma)$
and $\Lambda(\gamma)$ are stiff compared to those for $\tilde
s(\gamma)$ and $\chi(\gamma)$. Then we can safely replace Eqs.
(\ref{dotm4}) and (\ref{dotLambda4}) by their stationary solutions:
\begin{equation}
\label{m0}
m=m_0(\tilde s)=\cases{\tilde s/ |\tilde s| &if $|\tilde s|\le 1$\cr 1/\tilde s & if $|\tilde s| >1$}
\end{equation}
and
\begin{equation}
\Lambda = e^{-1/\chi}.
\end{equation}
Eqs. (\ref{sgamma}) and (\ref{dotchi4}) become:
\begin{equation}
\label{sgamma5}
{d \tilde s\over d\gamma}= \tilde\mu\,\left(1 - {2\,\epsilon_0\,\over q_0}\,e^{-1/\chi}\,q(\tilde s)\right),
\end{equation}
and
\begin{equation}
\label{dotchi5}
{d\chi\over d\gamma} = {2\,\epsilon_0\over c_0\,q_0}\,e^{-1/\chi}\,\tilde s\,q(\tilde s)\,(\chi_{\infty}-\chi),
\end{equation}
where
\begin{equation}
\label{q-sdef}
q(\tilde s) = \C C(\tilde s)\,\Bigl({\tilde s\over |\tilde s|}-m_0(\tilde s)\Bigr),
\end{equation}
and $q_0=\dot\gamma\,\tau_0$.

\section{First Fits to the DA Stress-Strain Curves}
\label{FirstFits}

\begin{figure}
\centering \epsfig{width=.5\textwidth,file=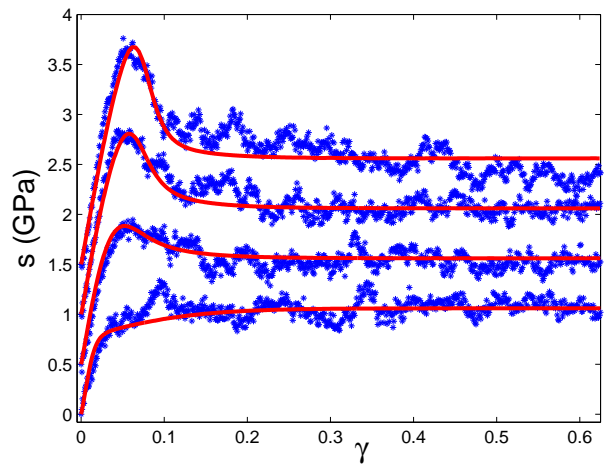} \caption{Theoretical stress-strain curves (solid lines) compared to the DA 
numerical simulation data \cite{DA1}. The parameters used are:
$\epsilon_0=1$, $s_f=1.06$GPa, $\mu=46$ GPa, $s_y=0.35$ GPa,
$c_0=0.18$ and $\chi_{\infty}=0.065$. The initial effective
temperatures are: $\chi_0$= $0.0559,\,0.0580,\,0.0612,\,0.0680$
(from top to bottom). The curves are shifted by 0.5 GPa for clarity.}
\label{best_fit}
\end{figure}

The first step in our STZ analysis of the DA data is to use Eqs.
(\ref{sgamma5}) - (\ref{q-sdef}) to fit the stress-strain curves in
Fig. 15(a) of \cite{DA1}, shown here in Fig. \ref{best_fit}.  
These two equations involve the parameters $q_0$ (the dimensionless 
strain rate), $s_y$ (needed in
order to convert from dimensionless stresses $\tilde s$ to measured
stresses in units GPa), $\tilde\mu$, $c_0$,
$\chi_{\infty}$, and the four initial values of the effective
temperature $\chi(0) = \chi_0$. 

We can obtain a certain amount of
information about these parameters directly from features of the
stress-strain curves before doing any
computation.  Note from Eqs. (\ref{m0}) and (\ref{q-sdef}) that no
plastic deformation occurs for $\tilde s <1$, no matter how large
$\chi$ might be.  Thus the lowest observed value of a stress that
marks a departure from elastic behavior is an upper bound for $s_y$.
The bottom stress-strain curve in Fig. \ref{best_fit}, which is the curve
with the largest initial value of $\phi$, seems to have a break
point that is about a factor of three less than the flow stress
$s_f$, which we see from the figure is about 1.06 GPa.  Therefore,
$s_y \cong s_f/3 \cong 0.35$ GPa, and $\tilde s_f \cong 3$. We take 
the shear modulus $\mu \cong 46$ GPa directly from the slope of 
the stress-strain curves in the elastic region; thus we choose 
$\tilde\mu \cong 130$. Setting the right-hand side of Eq.(\ref{sgamma5}) 
to zero, we find that the flow stress $\tilde s_f$ satisfies
\begin{equation}
\label{sf}
q_0 =\dot\gamma\,\tau_0= 2\,\epsilon_0\,e^{-1/\chi_{\infty}}\,q(\tilde s_f).
\end{equation}
Using this relation with the DA estimate of $\dot\gamma$ and the value of 
$\chi_{\infty}$ given below, 
we confirm that $\tau_0$ is of order femtoseconds as implied by 
the Stillinger-Weber interactions used in these simulations.

Once we have inserted Eq. (\ref{sf}) into Eqs. (\ref{sgamma5}) and
(\ref{dotchi5}), we can solve these equations and vary the remaining
parameters to fit the stress-strain curves in Fig. 15 (a) of
\cite{DA1}.  We find that the fits are disappointingly
insensitive to our choice of $\epsilon_0\,\exp(-1/\chi_{\infty})$ so
long as we stay within our constraint that this product be small.
However, our later analysis of the liquidlike fraction $\phi$ 
reveals that the  internally consistent value of
$\chi_{\infty}$ is $0.065$, which is slightly bigger than our {\it a
priori} order-of-magnitude estimate based on the experimental data, as shown in
\cite{JOHNSON} for example, but seems well within the accuracy
of our approximation for $\chi_{\infty}$ in \cite{04Lan} and the 
uncertainties of these small-scale simulations.  Therefore we have  chosen
$\epsilon_0=1$ and $\chi_{\infty}=0.065$.  Agreement between theory
and the numerical simulations then can be obtained for all four of
the stress-strain curves  by setting $c_0=0.18$ and adjusting only
$\chi_0$ for each curve.  Our best-fit results are shown in Fig.
\ref{best_fit}. The corresponding values of $\chi_0$ (from top to
bottom in the figure) are $0.0559,\,0.0580,\,0.0612,\,0.0680$. Note
that there is a gap in $\chi_0$ between the lowest three
values, for which the stress-strain curves are peaked, and the
largest, which shows no peak.

If it were useful to do so, we could improve the agreement 
between the simulated stress-strain
curves and our theory by making small adjustments of $s_y$ and $\tilde\mu$ for each
curve, consistent with the likelihood that the four relatively small
computational systems are not exactly comparable to each
other in their as-quenched states.  In the spirit of the discussion
to follow, however, we have chosen to assume that the four systems
reach effectively identical steady states after persistent shear
deformation, and to attribute the small discrepancies to the
statistical uncertainties visible in the data. The only systematic
feature of the stress-strain data that is not recovered by the
theory is the slow decrease of the flow stress with increasing
strain in the top curve (the one with the most pronounced stress
peak).  We believe that this behavior is caused by the emergence of
a nascent shear band as illustrated in Fig. 16(b) of \cite{DA1}.

\section{Second Thoughts}
\label{SecondThoughts}

Fitting the stress-strain curves, however, is only a part of the
challenge of interpreting the DA data \cite{DA1}. We also must
understand how the liquidlike fraction $\phi$ and the density $\rho$
relate to our STZ variables, especially the effective temperature
$\chi$.  Therefore, it is essential to know whether we can trust the 
values of $\chi_0$ deduced from the STZ analysis of the stress-strain 
curves. We argue in the following paragraphs that these values are 
not quantitatively reliable.

One of the key tenets of plasticity theory is that transient peaks 
observed in the stress-strain curves for well annealed samples 
occur because there is an initial lack of plasticity carriers 
in these systems, and new carriers must be generated by deformation 
before plastic flow can begin and the stress can relax.  
Demkowicz and Argon use the correlation between stress peaks and the 
liquidlike fraction $\phi$ to argue that $\phi$ is a direct 
measure of the population of plasticity carriers.  The STZ theory, 
as presently constituted, predicts stress peaks when -- and only 
when -- the initial STZ density is small. 

This tenet is not confirmed by the DA simulations.  Note first 
the data for $\phi$ as a function of strain $\gamma$ shown by 
DA in Fig.15 of \cite{DA1}, also shown here in Fig.\ref{phi-gamma}.  
There are four curves.  Two of them have
small initial values of $\phi(\gamma=0)= \phi_0$ and exhibit stress
peaks.  The corresponding values of $\phi(\gamma)$ rise monotonically 
to the steady-state value $\phi_{\infty} \cong 0.46$ as expected.  Another
curve starts with $\phi_0 > \phi_{\infty}$. The corresponding
stress-strain curve has no peak, and $\phi(\gamma)$ decreases to
$\phi_{\infty}$, again in accord with expectations.  In one case 
(the one with $\phi_0 \cong 0.46$), however, $\phi_0$ is slightly 
above $\phi_{\infty}$ but the stress still shows a peak.  Moreover, 
the corresponding density $\rho$ decreases with $\gamma$, implying 
that the denser liquidlike fraction is decreasing during the deformation.

This kind of behavior appears elsewhere in the DA papers 
\cite{04DA,DA1,DA2,06AD}.  It is clear in the constant-volume, MD
simulations, where one of the stress-strain curves has a peak, but
the corresponding $\phi(\gamma)$ remains nearly constant, and the
pressure increases instead of decreasing as it should if the
liquidlike fraction were growing.  And, as we note below, all of the PEM
simulations in \cite{DA2} show stress peaks, even the ones in which
$\phi_0$ is large and comparable to $\phi_{\infty}$. 

It seems to us that the most likely explanation for these discrepancies 
is that the ubiquitous stress peaks are artifacts of the step-strain 
simulations.  It is also quite possible, of course, that 
the STZ theory does not adequately account for the way in 
which stresses respond to rapid changes in the strain rate, and that the DA
simulations are simply out of range of our STZ analysis.  
Perhaps both explanations are in part correct. Nevertheless, we prefer the 
first explanation for the following reasons.

Demkowicz and Argon performed only MD simulations, and not PEM, 
at constant-pressure; but a number of features of their PEM results 
may be relevant here. They report that cascades of events were evident 
in their PEM simulations but were hard to detect in MD.  
Maloney and Lemaitre  \cite{ML04a,ML04b}, who
used only PEM, found that cascades were prevalent and sometimes so
large that they spanned their systems, which were only two
dimensional but larger in numbers of atoms than those used by DA.  
A related feature of the DA PEM simulations is 
that, even for rapidly quenched samples with large initial 
disorder, the stress-strain curves exhibit marked stress 
peaks followed by strain softening.  We suspect that these behaviors 
may be generally characteristic of step-strain processes, and 
correspondingly uncharacteristic of continuous strain mechanisms
that are more common in the laboratory. 

Note that, between each strain increment in 
PEM, the system has a probability of dropping into 
a low energy state from which it cannot escape without the 
application of a large force.  The large energy released 
when such a trapped state is destabilized may trigger 
cascades of smaller events. This trapping mechanism
may be especially important at the beginning 
of a shear deformation, because then the 
energy minimization starts with a disordered, as-quenched 
system.  As a result, the first energy drops may be large, 
and the initial stresses required to set the system into 
motion may be anomalously high.  Thus we expect transient 
stress peaks in PEM, even for initially disordered systems; 
and we expect large stress fluctuations even in steady-state 
conditions.  That is exactly what is seen by Demkowicz and Argon.

The difference between step strains and continuous shear has been
demonstrated recently in bubble raft experiments by Twardos and 
Dennin \cite{DENNIN05}.  Like Demkowicz and Argon, these authors 
subjected their strictly athermal system to both steady, slow 
shear and to discrete shear steps followed by relaxation periods 
long enough for most motion to cease. They monitored the stress 
during both procedures, with particular interest in the distribution 
of stress drops associated with irreversible plastic rearrangements.  
One of their most interesting results is that the average size of 
stress events decreases with decreasing shear rate for continuous 
strain, but increases for step strains.  That is, the stress relaxes
{\it via} a larger number of smaller drops for continuous deformation than
for step-strain motion.  Their interpretation of this result 
seems to us to be roughly consistent with our discussion of PEM simulations
in the preceding paragraph; but neither they nor we claim a full 
understanding of this phenomenon 

In their MD simulations, Demkowicz and Argon use step strains 
as opposed to a continuous strain rate.  Apparently, allowing the system
to relax for a time at $T = 300\,K$ between strain steps removes the 
effects of at least some of whatever structural irregularities are producing 
cascades and stress peaks in the PEM simulations.  It seems likely, however, 
that some elements of the athermal PEM-like, step-strain behavior 
persist in the MD results.  Our hypothetical low-energy trapping 
states may be less likely to be sampled in step-strain MD, in 
which case the initial transients and steady-state 
stress fluctuations might be smoother, as indeed they are. 
But the step-strain MD procedure is not the same as a 
continuous one in which the strain is incremented on the  
time scale at which the molecular motions are resolved.  In the 
DA simulations, the strain is incremented only once in about 
ten or more atomic vibration periods.
We see little reason to believe that these two simulation procedures will 
produce precisely the same responses to rapidly changing driving
conditions.  

This situation forces us to conclude -- reluctantly in
view of the quality of the apparent agreement between the
simulations and STZ theory shown in Fig. \ref{best_fit} -- that the
values of the $\chi_0$'s stated above and in the caption to Fig.
\ref{best_fit} cannot be trusted.

\begin{figure}
\centering \epsfig{width=.48\textwidth,file=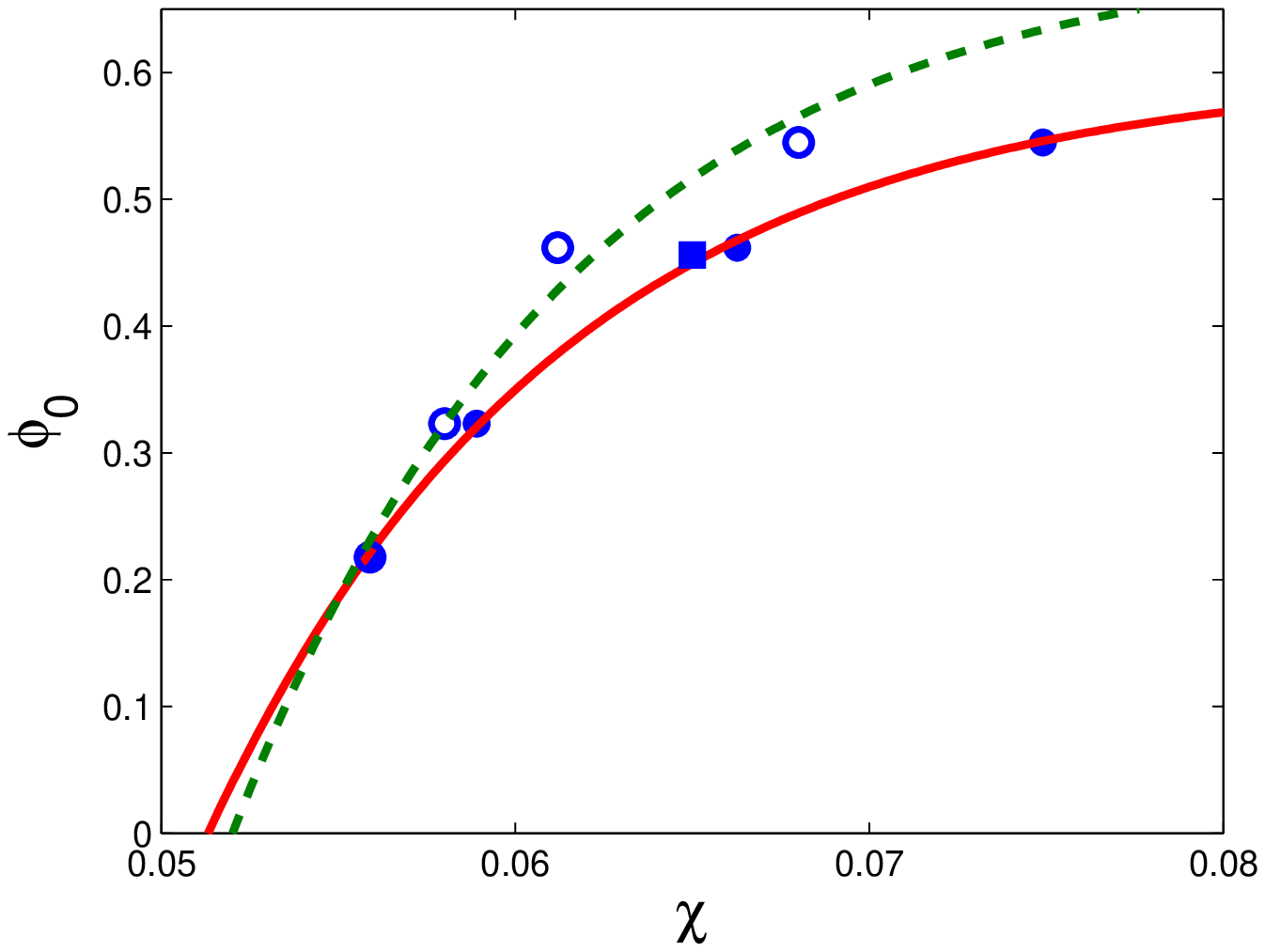} 
\centering \epsfig{width=.48\textwidth,file=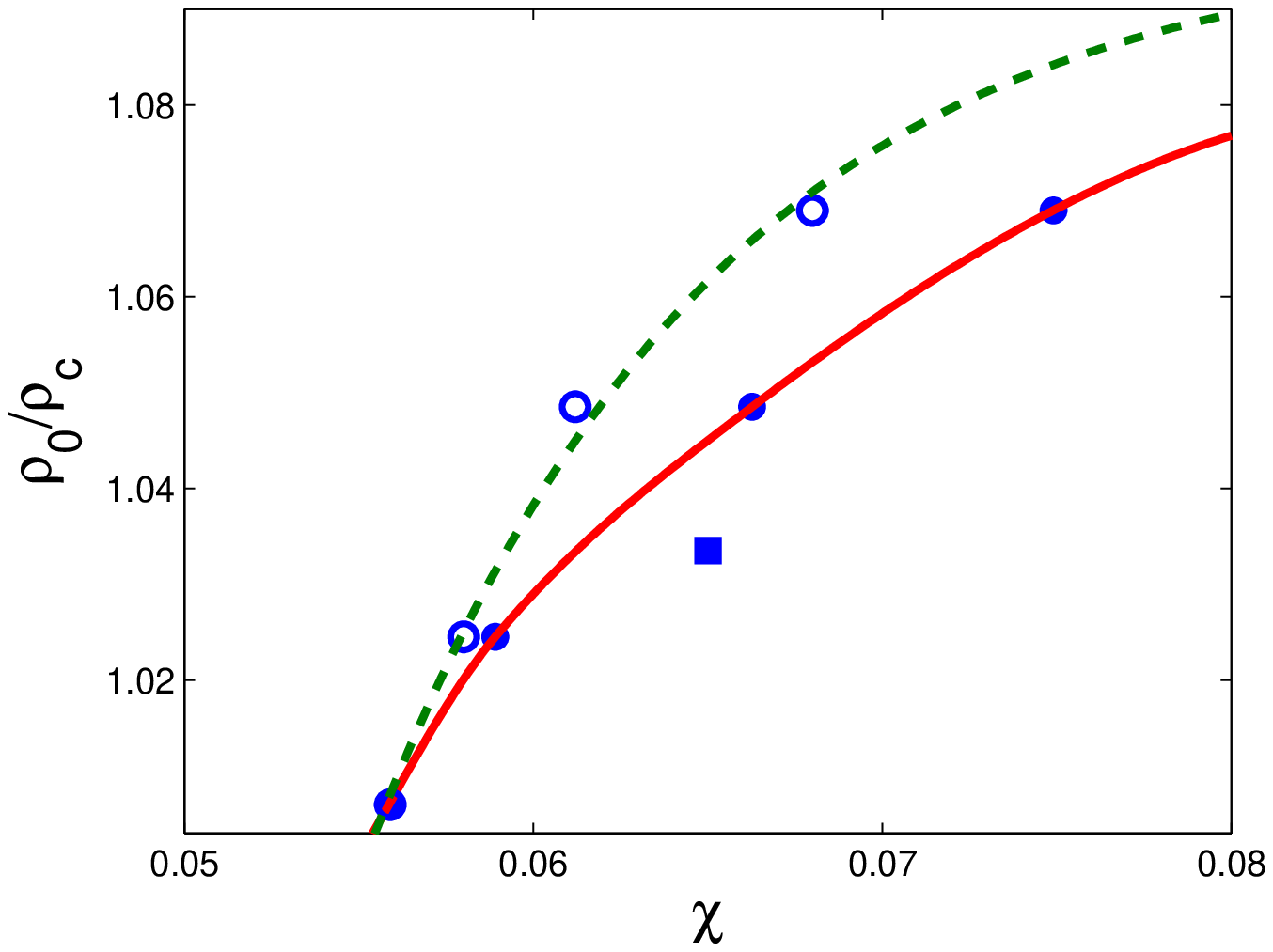}
\caption{Upper panel: The equation of state
$\phi=\phi_0(\chi)$ obtained by fitting the four data points (filled
circles) plus $\phi_{\infty}=\phi_0(\chi_{\infty})=0.458$ 
(filled square). The $\phi_0$'s are the DA numerical
simulation data \cite{DA1}; the adjusted $\chi_0$'s are 
$0.0559,\,0.0589,\,0.0663$, and $0.0749$, with 
$\chi_{\infty} = 0.065$. The solid line is
our fit to these points in Eq. (\ref{EOS_phi}).  The open circles 
are the observed $\phi_0$'s plotted 
against the unadjusted, best-fit, $\chi_0$'s. The dashed line is 
intended only as a guide for the eye. Lower panel: The equation of
state $\rho=\rho_0(\chi)$ obtained by a procedure similar to the one
described above except that the point $\rho_{\infty}=1.0335$ 
(filled square) is not included in the fit. We attribute its 
displacement from the equation of state to a stress-induced dilation.
The density is normalized by $\rho_c$, the
density of crystalline silicon (diamond cubic). The solid line
is our smooth fit to the data.  Again, the open circles and and dashed line indicate the unadjusted equation of state.}  \label{EOS}
\end{figure}

\section{Quasithermodynamic Hypotheses}
\label{Quasithermodynamics}

Our inability to deduce the $\chi_0$'s from the stress-strain 
data has led us to probe more deeply into the meaning of the 
effective temperature in glass dynamics.  We base our analysis on 
a set of quasithermodynamic hypotheses in which we assume that the 
steady-state properties of the configurational degrees of freedom 
below the glass transition are determined by the effective temperature
$T_{eff}= (E_{STZ}/k_B)\,\chi$, in close analogy to the way in which they
are determined by the bath temperature $T$ above that transition.  
In true thermodynamic equilibrium, extensive quantities such as the density 
$\rho$, the internal energy $U$, or the liquidlike fraction 
$\phi$ in the DA simulations
are functions of $T$ and the pressure $P$.  In other words, these quantities
obey equations of state.  Below the glass transition, the configurational 
degrees of freedom -- that is, the positions of the atoms in their 
inherent states -- fall out of equilibrium with $T$ because thermally 
activated rearrangements are exceedingly slow or impossible.  The most 
probable configurations in such situations maximize an entropy, thus 
the statistical distribution of these configurations is Gibbsian 
with $T=T_{eff}$.   

Accordingly, our first hypothesis is that the high-$T$ equilibrium 
equations of state are preserved in the glassy state as expressions 
for the configurational parts of the extensive quantities $\rho$, 
$U$, $\phi$, {\it etc.} in terms of the intensive quantities 
$T_{eff}$, $P$, and the shear 
stress $s$.  In particular, we propose  that the effective 
temperatures $\chi_0$, multiplied by $E_{STZ}/k_B$, are the bath 
temperatures $T_0$ at which the DA simulation samples fell out of thermal 
equilibrium during cooling, and that the configurational parts of $\rho$, 
$U$, $\phi$, {\it etc.} were fixed at those temperatures. Our 
specialization to configurational parts recognizes 
that, for example, $\rho$ undergoes ordinary thermal expansion 
at small $T$ and that the total internal energy $U$ 
includes the kinetic energy.  Both of those non-configurational parts
are uninteresting for present purposes.  Since we consider only 
$P=0$ situations and work only at fixed $T$, we omit explicit 
dependences on those variables. Shear dilation may be relevant, 
especially for determining $\rho$; but so long as we are considering 
only pre-deformation properties with $s=0$, we also omit explicit $s$ 
dependence. We return later to the question of shear dilation, and 
we also postpone a discussion of the potential energy.  For the moment, 
we write simply: $\phi = \phi_0(\chi)$, $\rho = \rho_0(\chi)$.  

As a first test of this hypothesis, we show by the open circles and 
dashed lines in both 
panels of Fig. \ref{EOS} the functions $\phi_0(\chi)$ and $\rho_0(\chi)$ 
obtained with the measured values of $\phi_0$ and $\rho_0$ and our best-fit 
values of $\chi_0$ from Fig. \ref{best_fit}.  We do, in fact, 
find qualitatively plausible behavior.  With our uncertainties about 
the $\chi_0$'s, however, we are obliged to look harder and bring other 
considerations to bear on this analysis.
\begin{figure}
\centering \epsfig{width=.5\textwidth,file=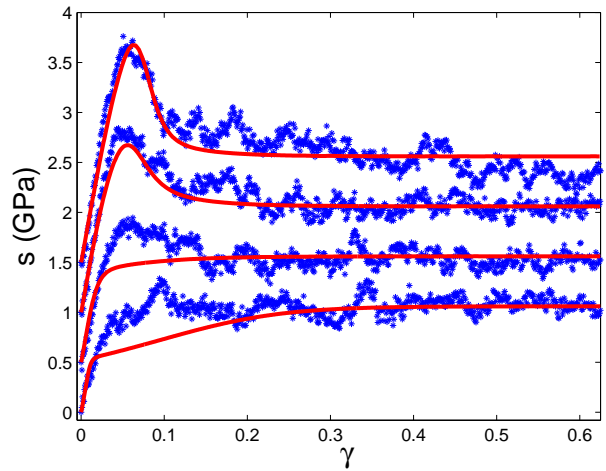} \centering
 \caption{Modified stress-strain curves using $\chi_0=0.0559,\,0.0589,\,0.0663$, and $0.0749$.
 All other parameters are the same as those used in Fig. \ref{best_fit}.}
\label{modified}
\end{figure}

We know some qualitative features of these equations of state.
First, we are dealing with a glass transition, and therefore we
expect that $\phi_0(\chi)$ should extrapolate to some fixed value, 
possibly zero, at a nonzero Kauzmann temperature.  In the following
discussion, we  define $\chi_K$ to be the effective temperature at
which the extrapolated $\phi_0(\chi)$ vanishes; but we make no
assumption about whether this is actually a Kauzmann temperature, 
or whether an ideal glass transition actually
occurs at this or any other point.  

Another feature is suggested
by the binary, solidlike/liquidlike, nature of the atomic structure
described by Demkowicz and Argon.  If this were just a simple binary
mixture, and if the energy of the mixture were proportional just to
the density of the higher energy, liquidlike component, then the
entropy would be a maximum at $\phi = 1/2$ and the temperature would
diverge at that point.  Obviously, amorphous silicon is not so
simple.  Nevertheless, the observed $\phi$ does seem to have an
upper bound not too far above $\phi = 1/2$;  thus, we expect
that $\phi_0(\chi)$ saturates at some value $\phi_0 =\phi_{{\rm lim}}\ge 0.55$ for
large $\chi$.

Yet another consideration is that, in principle, we ought to be able 
to deduce the $T_0$'s directly from the graphs of
density {\it versus} temperature for different quench rates shown in Fig.
1 of \cite{DA1}.  The slowest quench is the closest to full
equilibrium, and the points at which the faster quenches fall away
from it are the three higher $T_0$'s.  The lowest $T_0$ must be near 
the inflection point on the slow-quench curve.  Unfortunately, the 
best we can do with the data at hand is estimate the $T_0$'s  
to about $\pm 20\,K$, and even then we must be cautious about 
systematic uncertainties in the simulations.  Nevertheless, we can use
this process as a rough check on the modifications of the $\chi_0$'s 
that we propose below.

We come now to our second and more speculative 
hypothesis -- that the point
$(\chi_{\infty},\,\phi_{\infty})$ ought to lie on the equilibrium
equation-of-state curve; that is, $\phi_{\infty} =
\phi_0(\chi_{\infty})$. Here we are assuming that persistent shear 
deformation in an athermal system rearranges the atomic 
configurations in a way that is statistically equivalent to 
thermally driven rearrangement,
except that the relevant temperature is the effective disorder
temperature $T_{eff}$ instead of the bath temperature $T$. We 
also are assuming that $\phi_{\infty}$, being a ratio of two 
populations, is insensitive to changes in the volume of the 
system that might occur during constant-zero-pressure simulations.  
In contrast, $\rho_{\infty}$ would decrease if, for example, 
the system undergoes a shear-induced dilation.

With these considerations in mind, our next step 
is to refine the estimates of the
$T_0$'s by requiring that they produce smooth equations of state
with the qualitative features hypothesized above, and that the 
observed $\phi_{\infty}\cong 0.458$ be equal to 
$\phi_0(\chi_{\infty})$. Accordingly, we have computed $\phi =\phi_0(\chi)$ 
and $\rho=\rho_0(\chi)$ by adjusting the $\chi_0$'s so as to optimize
agreement with all the previously discussed constraints. Our results
are shown by the solid lines in Fig. \ref{EOS}.  As expected, 
$\rho_{\infty} < \rho_0(\chi_{\infty})$, because the flow stress 
$\tilde s_f$ at that point seems easily large enough to cause dilation.  
(Throughout this paper, 
values of $\rho$ are in units of $\rho_c= 2323.8\, {\rm kg/m}^3$, the 
density of crystalline, diamond-cubic, silicon.) The effective Kauzmann
temperature is $\chi_K=0.051$, and the upper bound for $\phi$ is
$\phi_{{\rm lim}}=\phi(\chi\!\to\!\infty)\cong 0.6$. The smooth curve
that we have fit to the equation of state for $\phi$, shown in the
upper panel of Fig. \ref{EOS}, is
\begin{equation}
\label{EOS_phi}
\phi_0(\chi)=\phi_{{\rm lim}}\,\left(1-e^{-a\,(\chi-\chi_K)}\right),
\end{equation}
where $a = 100$. Other important parameters are the four $T_0$'s:
$1100\,K,\,1160\,K,\,1305\,K,\,1475\,K$, which are consistent with 
our crude direct estimates from Fig. 1 in \cite{DA1}.  Note that 
the large gap between the lowest three unadjusted $T_0$'s and the 
upper one has disappeared, and that the adjusted points fit on a 
smoother curve than the unadjusted ones. We also find that
$T_{\infty}=1280\,K$. Then, with $\chi_{\infty} = 0.065$, we obtain
$E_{STZ}/k_B\cong 21,000\,K$, or about $1.3$ ev. The associated
values of $\chi_0$ are $0.0559,\,0.0589,\,0.0663$, and $0.0749$.  

Using these adjusted values of the $\chi_0$'s, 
we have recomputed and plotted the stress-strain curves in Fig.
\ref{modified}, shown again in comparison with the simulation
data. As expected, the non-STZ peak has disappeared, and there 
is more rounding in the bottom, most rapidly quenched case, but the 
stress peaks for the two most deeply quenched systems remain 
almost unchanged. 

Finally, within the context of quasithermodynamic equilibrium, 
we note that a relation between the potential energy and $\chi$ 
has been proposed and confirmed by 
Shi {\it et al.}.\cite{SHI06} These authors describe molecular-dynamics 
simulations of shear banding in two-dimensional, non-crystalline, 
Lennard-Jones mixtures.  Their analysis of these simulations goes 
beyond our own in at least one important way.  They note 
that the plastic strain rate predicted by the STZ theory, as well as by 
other flow-defect theories, has the form $\exp(-1/\chi)$ multiplied by a 
function of the shear stress, essentially our factor $q(\tilde s)$ defined 
in Eq.(\ref{q-sdef}).  Force balance requires that $\tilde s$ be a constant 
across the shear band, thus a measurement of the position-dependent strain 
rate is a measure of the position dependence of $\chi$.  Shi {\it et al.} 
then postulate that the potential energy depends 
linearly on $\chi$.  They compute the position-dependent potential energy 
directly from their simulation data and find that it maps accurately onto 
$\chi$ as predicted by the spatially varying strain rate. The 
importance of this observation is that they are 
looking at a weakly nonequilibrium situation in which the potential energy and 
$\chi$ are varying continuously in space, but only very slowly in time, 
along the postulated linear equation of state. Their system remains 
in quasithermodynamic equilibrium during this variation because, unlike 
the DA simulations discussed here, it is not undergoing a fast transient 
response to a sudden change in the applied stress. Thus their result
anticipates and fits accurately into our quasithermodynamic picture.

\section{Departures from Quasithermodynamic Equilibrium}
\label{Departures}

\begin{figure}
\centering \epsfig{width=.5\textwidth,file=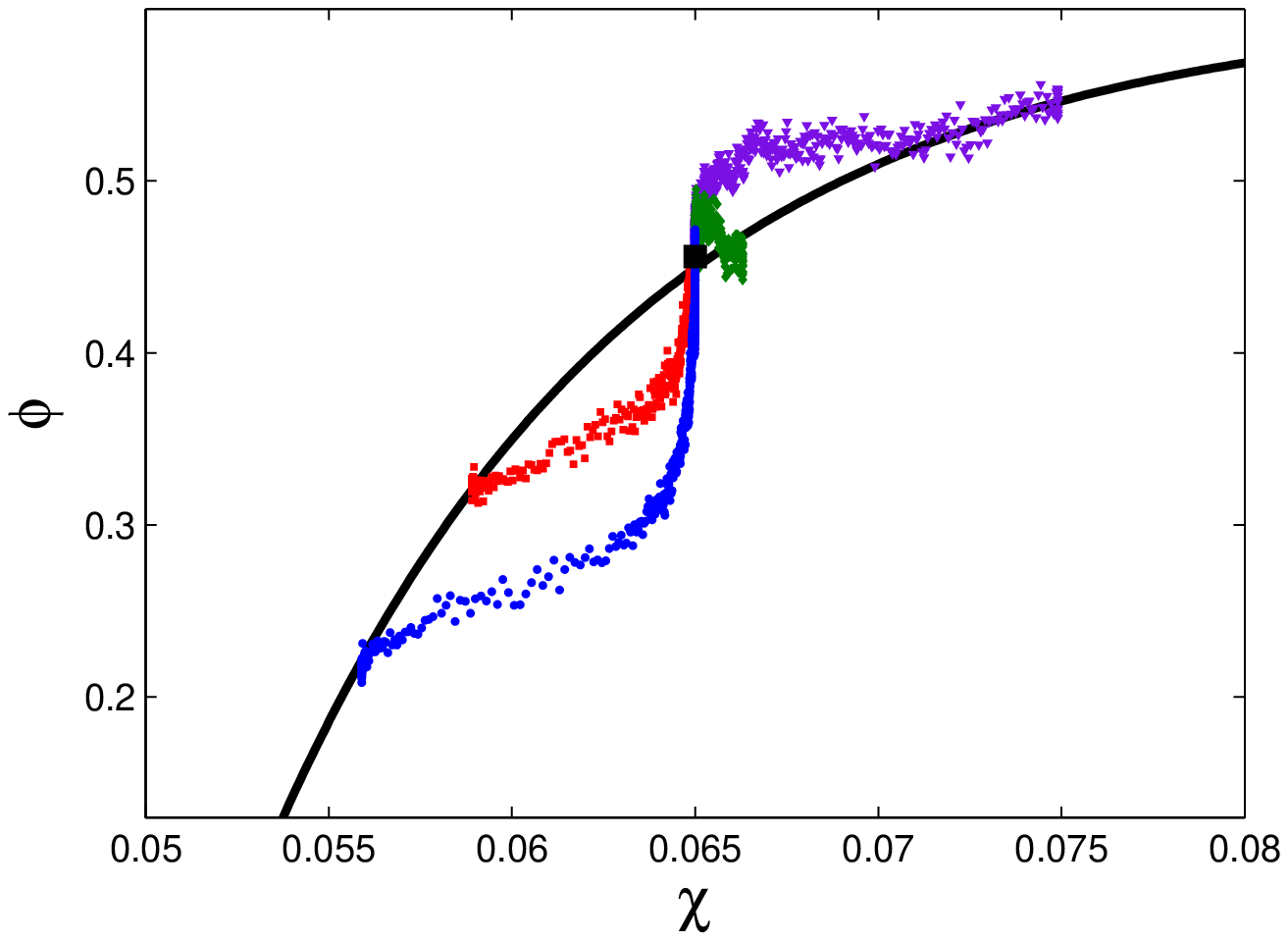} \centering
\epsfig{width=.5\textwidth,file=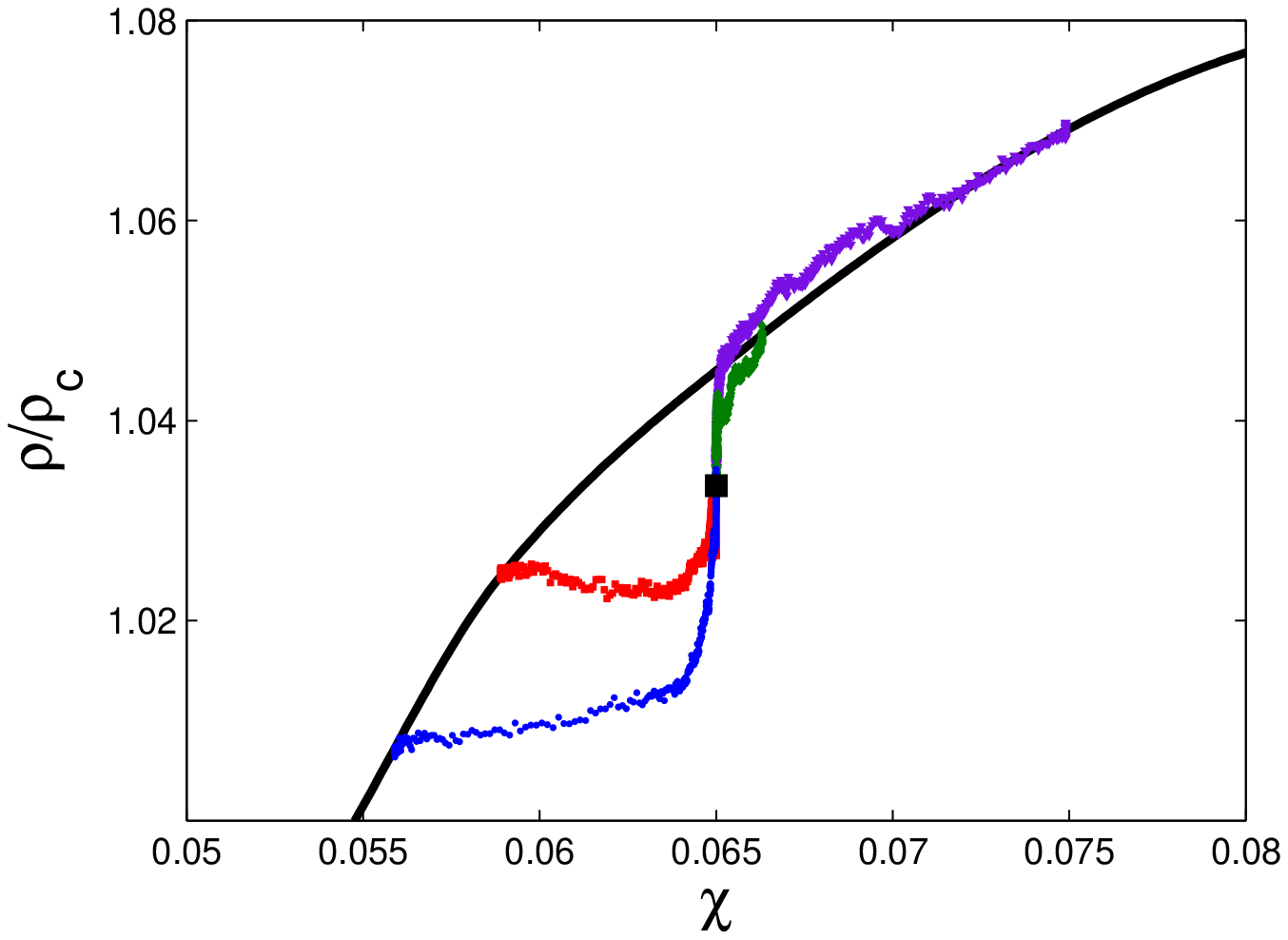} \caption{Upper panel: The
DA numerical simulation data for $\phi$ versus the theoretical $\chi$
for the four cases considered in Ref. \cite{DA1}. The equation of
state $\phi=\phi_0(\chi)$ was added to stress the nonequilibrium nature of the 
deformation-induced dynamics of $\phi$.  Lower panel: The corresponding plot 
for $\rho/\rho_c$. The points $(\chi_{\infty},\,\phi_{\infty})$ and 
$(\chi_{\infty},\,\rho_{\infty})$ are marked by solid squares.}
\label{noneq}
\end{figure}

The next question is whether the equilibrium equations of state 
shown in Fig. \ref{EOS} are obeyed
under nonequilibrium conditions during deformation. Apparently they
are not.  To see this, in Fig. \ref{noneq}, we have added to our
equation-of-state graphs four data sets showing the observed values
of $\phi$ and $\rho$ as functions of our calculated values of $\chi$ for the
four different deformation histories. In all cases, the observed
values of $\phi$ and $\rho$ initially fall off the equilibrium
curves and do not come together again until they approach $\chi_{\infty}$.
This nonequilibrium behavior can be seen more directly by comparing
the three panels of Fig. 15 in \cite{DA1}.  It is clear there that the
rate at which the stress relaxes to the flow stress is faster than
the rates at which either $\phi$ or $\rho$ relax to
their steady-state values. 

To account for the transient, nonequilibrium behavior of $\phi$, we 
propose an equation of motion analogous to Eq. (17) in
\cite{06AD}, but with a form similar to our Eq. (\ref{dotchi5}):
\begin {equation}
\label{dotphi}
{d\phi\over d\gamma} = {2\,\epsilon_0\over c_1\,q_0}\,e^{-1/\chi}\,
\tilde s\,q(\tilde s)\,\left(\phi_0(\chi)-\phi\right),
\end{equation}
where $c_1$ is a constant similar to $c_0$.  The idea here is that
the disorder described by $\chi$ may rise at its own rate, but the
liquidlike-solidlike  reorganization associated with $\phi$ may not
catch up instantaneously.  We assume that the underlying mechanism
that determines these rates is still the rate of energy dissipation.
There is no other comparably simple and basic coupling between
mechanical deformation and the internal degrees of freedom that
satisfies the requirement that it be a non-negative scalar. Note that 
Eq.(\ref{dotphi}) is supplementary to the STZ equations of motion, 
Eqs.(\ref{sgamma5}) and (\ref{dotchi5}); it assumes that the intensive
quantities $\tilde s(\gamma)$ and $\chi(\gamma)$ are unchanged from the 
STZ predictions and that they continue to control the behavior 
of $\phi$ even away from equilibrium.  This is a 
strong assumption, especially near the initial stress transients 
where we already know that there is some mismatch between the STZ 
theory and the DA simulations.

\begin{figure}
\centering \epsfig{width=.5\textwidth,file=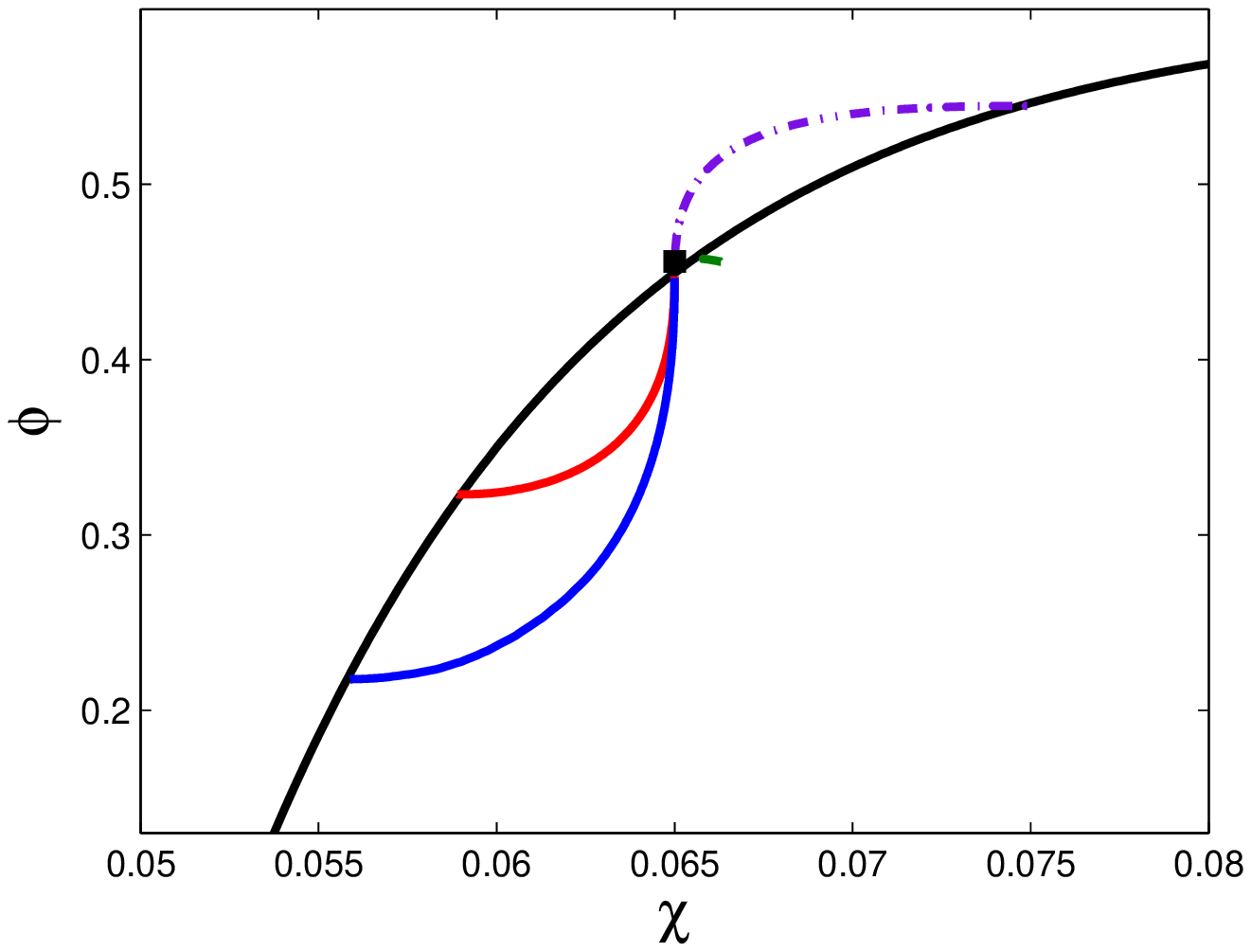} \centering \epsfig{width=.5\textwidth,file=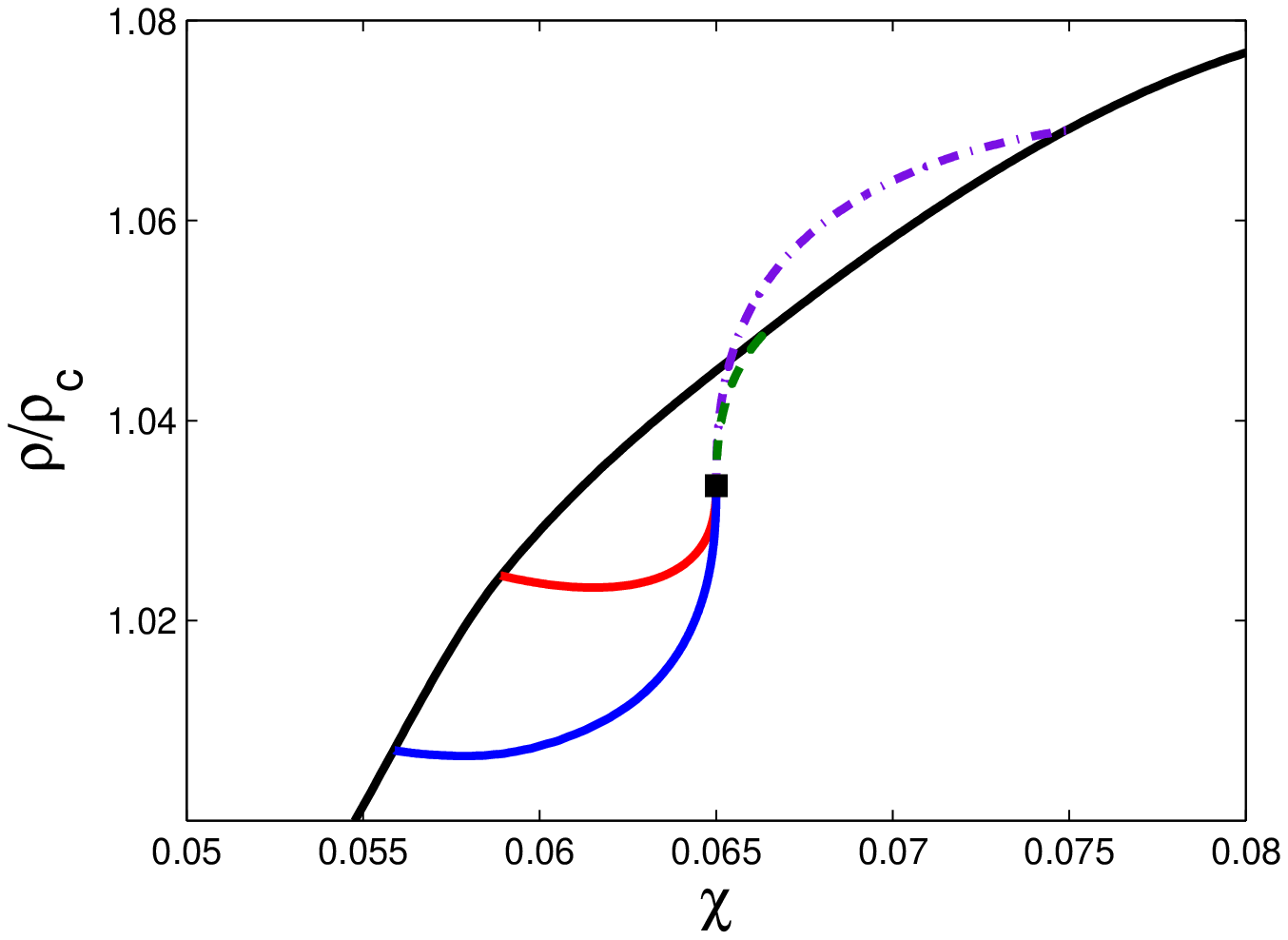}\caption{Upper panel:$\phi$
as a function of $\chi$ using the nonequilibrium Eq. (\ref{dotphi})
with $c_1=0.42$. The equilibrium equation of state was added for illustration.
The curves should be compared to the DA data shown in the upper panel
of Fig. \ref{noneq}. Lower panel: $\rho$ as a function of $\chi$ using 
the nonequilibrium Eq. (\ref{dotrho}) with $c_1=0.42$. These curves should be 
compared to the DA data shown in the lower panel of Fig. \ref{noneq}. 
The points $(\chi_{\infty},\,\phi_{\infty})$ and 
$(\chi_{\infty},\,\rho_{\infty})$ are marked by solid squares.} \label{noneq-theory}
\end{figure}

\begin{figure}
\centering \epsfig{width=.5\textwidth,file=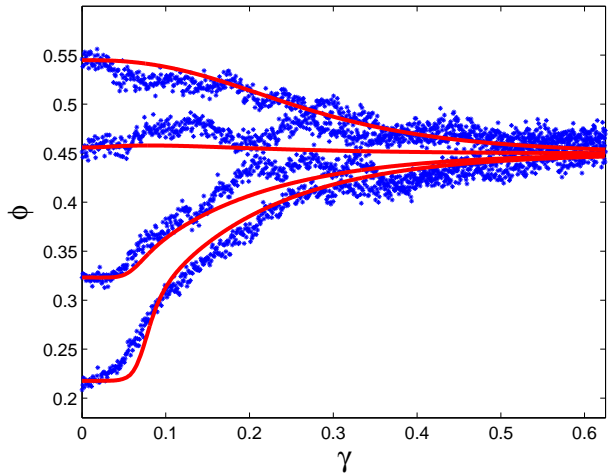}
\caption{Theoretical predictions of $\phi(\gamma)$  for all four quenches, based on  Eq. (\ref{dotphi}) with $c_1=0.42$, compared with the DA simulation data.  } 
\label{phi-gamma}
\end{figure}

Our results for the four functions $\phi(\chi)$, determined using 
Eq.(\ref{dotphi}), are shown in Fig.\ref{noneq-theory}.  The 
corresponding functions $\phi(\gamma)$ are shown in Fig.\ref{phi-gamma}, 
here in comparison with the DA data.  The only adjustable parameter 
is $c_1$, which we choose to be $0.42$.  The agreement seems to be 
within the uncertainties of the data.

\begin{figure}
\centering \epsfig{width=.5\textwidth,file=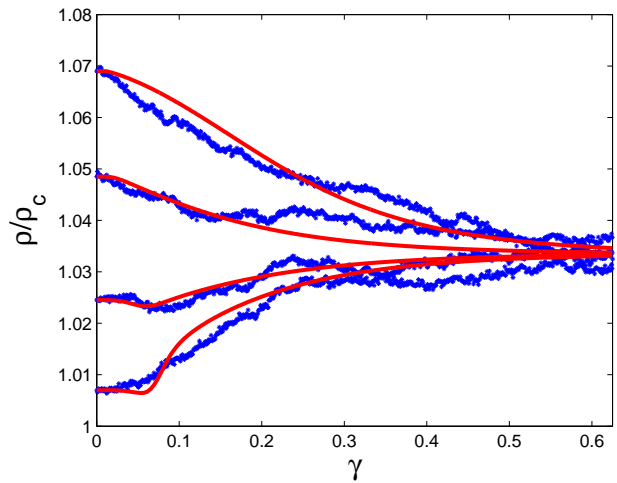}
\caption{Theoretical predictions of $\rho(\gamma)$  for all four quenches, based on  Eq. (\ref{dotrho}) with $c_1=0.42$, compared with the DA simulation data.  } 
 \label{rho-gamma}
\end{figure}

We continue our development of the nonequilibrium quasithermodynamic theory 
by writing an equation analogous to Eq. (\ref{dotphi}) for 
the function $\rho(\gamma)$:
\begin{equation}
\label{dotrho}
{d\rho\over d\gamma} = {2\,\epsilon_0\over c_1\,q_0}\,
e^{-1/\chi}\,\tilde s\,q(\tilde s)\,\left(\rho_0(\chi)-\rho-\delta\rho_0\right),
\end{equation}
Here, $\delta\rho_0\cong 0.012$ is the dilation-induced shift of the equilibrium 
density shown by the displacement of $\rho_{\infty}$ from the 
equilibrium curve $\rho_0(\chi)$ in the lower panel of Fig.\ref{EOS}.  
To integrate Eq.(\ref{dotrho}), we have approximated $\rho_0(\chi)$ by a smooth 
polynomial. The factor $c_1=0.42$ needed to fit the density data 
is the same as in Eq. (\ref{dotphi}).  The corresponding graphs of 
$\rho(\gamma)$, along with the DA data, are shown in Fig. \ref{rho-gamma}.  
Again, the agreement seems to be within the uncertainties; but here we are
pushing the theory too far for comfort.  Our quasithermodynamic hypotheses
imply that $\rho_0(\chi)$ ought to be a function of the shear stress 
$\tilde s$ as well as $\chi$, and that the dilation approximated here by
$\delta\rho_0$ should be part of that generalized equation of state.  
In particular, $\delta\rho_0$ should be the dilational change in the 
density when $\tilde s$ is equal to the flow stress $\tilde s_f$. We
have in fact tried $\delta\rho_0 \propto \tilde s^2$ (as in nonlinear 
elasticity), but the results are distinctly unsatisfactory at small 
$\gamma$ where neither Eq.(\ref{dotrho}) nor the STZ theory itself may
accurately describe the fast transient.  Equations (\ref{dotphi}) 
and (\ref{dotrho}) do account for the relatively slow relaxation of 
$\phi(\gamma)$ and $\rho(\gamma)$ as compared to that of 
$\tilde s(\gamma)$. This level of success seems to be as much as we 
can expect from the theory at this stage in its development.

\section{Summary and Concluding Remarks}
\label{Conclusions}

The unique aspect of the Demkowicz and Argon simulations is their 
measurement of extensive quantities -- the mass density $\rho$ and the 
liquidlike fraction $\phi$ -- in parallel with conventional stress-strain 
curves.  They argue convincingly and importantly that $\phi$ is closely 
related to the density of plasticity carriers.  However, their remarkably 
complete report includes some results that make it seem that this relationship 
may be neither simple nor direct.  The main purpose of our investigation has 
been to learn more about that relationship.

Our proposed interpretation of the DA simulations is based on the athermal STZ 
theory \cite{BLP06}, which 
we believe captures the central features of amorphous plasticity 
-- at least for processes that are not too rapidly varying as functions
of space or time. In order to discuss quantities like $\phi$, however, 
we have had to go beyond the STZ theory.  We have hypothesized that, 
in low-temperature, steady-state, nonequilibrium conditions, 
$\rho$ and $\phi$ and presumably other such quantities 
are related to the effective disorder temperature $\chi$ {\it via} 
quasithermodynamic equations of state. With this hypothesis, we have 
developed a theoretical interpretation of the DA 
simulations that seems physically satisfying, internally self-consistent, 
but interestingly incomplete. If confirmed by further tests, this 
quasithermodynamic theory could become a useful tool for predicting the 
nonequilibrium mechanical behavior of amorphous solids. 

There are many open issues.  Perhaps the most urgent of these is the
interpretation of the DA step-strain simulation technique.  
The DA data do show that a well 
annealed sample with an initially small $\phi$ exhibits a transient peak 
in its stress-strain curve.  However, the converse seems not necessarily 
to be true.  Stress peaks sometimes appear in the DA results in cases where
the initial $\phi$ is large and does not increase during plastic deformation,
a behavior that is inconsistent with the presumed relation between $\phi$ 
and the density of plasticity carriers.  After examining other possibilities 
and looking at other examples of step-strain procedures 
(see Sec.\ref{SecondThoughts}), we have concluded that the stress peaks 
observed in cases with large initial $\phi$ are most likely artifacts of the 
numerical step-strain procedure.  
Once we allowed ourselves that flexibility and found alternative ways of 
estimating the initial effective temperatures for those cases, we found that our quasithermodynamic picture fits together quite well.

In our opinion, one of the most important next steps in determining the 
limits of validity of the quasithermodynamic STZ theory would be 
to redo the DA analysis, including the measurement of the liquidlike 
fraction $\phi$, with continuous strain MD.  
So far as we know, nobody before Demkowicz and Argon has ever 
done anything like this -- making independent, simultaneous measurements 
of the density of plasticity carriers and the stress-strain response for 
differently quenched systems.  Discovering the relation between 
continuous and step-strain simulations in this context seems likely 
to be extremely useful at least for understanding 
the numerical simulations.  It could also be a big step forward 
in understanding amorphous plasticity.

Another obviously open issue is the validity of the STZ theory in situations
where the system is responding to rapid changes in loading conditions, as in
the DA numerical simulations where the strain rate is large and is turned on
instantaneously both at the beginning of the process and at each strain step. 
The present version of STZ theory \cite{BLP06} is a mean-field approximation in
which interactions between zones are included only on average, and fluctuations
are neglected.  It is possible that neither the STZ theory as 
presently formulated, nor the conventional explanation of 
stress peaks in terms of plasticity carriers, can fully account 
for stress transients during rapid changes in loading conditions, even 
in continuous-strain MD simulations.  From this point of view, a 
continuous-strain version of the DA simulations seems doubly important.  
If new simulations showed no qualitative discrepancy between continuous 
and step-strain procedures, we would know that the STZ theory is missing 
some essential ingredients. We might then try to include correlations and 
fluctuations by developing a more detailed statistical theory of STZ's 
with varying thresholds, perhaps with noisy interactions between them 
as proposed recently by Lemaitre and Caroli \cite{LC06}.  

In a similar vein, we must ask about the limits of validity of our 
quasithermodynamic hypotheses.  It will be especially important to look harder at 
Eqs.(\ref{dotphi}) and (\ref{dotrho}), which determine the rates at which 
$\phi$ and $\rho$ relax to their quasithermodynamic equilibrium values. 
Our quasithermodynamic picture is based on the assumption that 
the effective temperature $\chi$ is the intensive variable -- 
the thermodynamic force -- that drives the extensive quantities 
$\phi$, $\rho$, and the STZ variables $\Lambda$ and $m$.  According 
to our equations of motion, $\Lambda$ and $m$ are tightly slaved to $\chi$.  
We have postulated equations of state relating $\phi$ 
and $\rho$ to $\chi$ under equilibrium or steady-state conditions, 
and have further proposed in Eqs.(\ref{dotphi}) and (\ref{dotrho}), 
on what seem to us to be general grounds, 
that $\phi$ and $\rho$ are more loosely slaved to $\chi$ than 
are $\Lambda$ or $m$ during excursions from equilibrium.  Here, as 
in the questions regarding the stress response, we need to learn 
whether the equations of motion for $\phi$ and $\rho$ are accurate when 
those excursions from equilibrium are faster, say, than the relaxation 
rate of $\chi$.  

In short, we are asking how far this theory can be pushed.
Can it, for example, always be used to predict stress-strain transients 
under realistic experimental conditions, where strain rates 
are very much smaller than those in MD simulations? Or does it 
generally become inaccurate near stress peaks?  Can it be
used to predict plastic deformation near the tip of 
an advancing crack?  More generally: Is the effective temperature 
$\chi$ really such a dominant state variable? What other internal 
variables might become relevant for describing fast processes? What 
real or computational experiments might help to answer 
such questions?

\begin{acknowledgments}
We thank M.J. Demkowicz and A.S. Argon for providing the data on 
which this research was based and for permission to use it here. 
J.S.L. also thanks C. Caroli, M. Falk, A. Lemaitre, and C. Maloney for 
important ideas and information. 
E. Bouchbinder was supported by a doctoral fellowship from 
the Horowitz Complexity Foundation. J.S. Langer was supported 
by U.S. Department of Energy Grant No. DE-FG03-99ER45762. 
I. Procaccia acknowledges the partial financial support of 
the Israeli Science Foundation, the Minerva Foundation, 
Munich, Germany, and the German-Israeli Foundation. 
\end{acknowledgments}

\end{document}